\documentclass{emulateapj}

\shorttitle{Nonthermal properties of SNR G1.9+0.3}
\shortauthors{Ksenofontov et al.}

\newcommand{\gsim}{\,\raisebox{0.2em}{$>$}\!\!\!\!\!
\raisebox{-0.25em}{$\sim$}\,}

\newcommand{\gr}{$\gamma$-ray \,}
\newcommand{\grs}{$\gamma$-rays \,}

\begin{document}

\title{Nonthermal properties of supernova remnant G1.9+0.3}

\author{
L.T.~Ksenofontov\altaffilmark{1},
H.J.~V\"olk\altaffilmark{2},
and E.G.~Berezhko\altaffilmark{1}
}

\altaffiltext{1}{Yu.G. Shafer Institute of Cosmophysical Research and Aeronomy,
                     31 Lenin Ave., 677980 Yakutsk, Russia}
\altaffiltext{2}{Max Planck Institut f\"ur Kernphysik,
                Postfach 103980, D-69029 Heidelberg, Germany}

\email{ksenofon@ikfia.ysn.ru}

\begin{abstract}
The properties of the -- presumably -- youngest Galactic supernova remnant
(SNR) G1.9+0.3 are investigated within the framework of nonlinear kinetic
theory of cosmic ray acceleration in SNRs. The observed angular size and
expansion speed as well as the radio and X-ray emission measurements are
used to determine relevant physical parameters of this SNR. Under the
assumption that SNR G1.9+0.3 is the result of a Type Ia supernova near the
Galactic center (at the distance $d=8.5$~kpc) the nonthermal properties are
calculated. In particular, the expected TeV gamma-ray spectral energy
density is predicted to be as low as $\epsilon_{\gamma}F_{\gamma} \approx
5\times10^{-15}$~erg~cm$^{-2}$~s$^{-1}$, strongly dependent
($F_{\gamma}\propto d^{-11}$) upon the source distance $d$.
\end{abstract}


\section{Introduction}
G1.9+0.3 has been known as a potentially young shell type Galactic supernova
remnant (SNR) of very small angular size \citep{gg84}. Recently, the 
interest in
this SNR was revived by \citet{rey08, rey09} who analyzed the expansion rate of
the object and deduced an age $t_\mathrm{SN}$ of about
100~yr, which makes it the youngest known SNR in the Galaxy. Although the
expansion rate was derived by a comparison of radio observations in 1985 and
{\em Chandra} observations in 2007, this rate has been confirmed very soon
thereafter by independent radio observations \citep{green08, murphy08}.

According to \citet{rey08}, the line-free X-ray emission has a pure
synchrotron
origin which clearly indicates that effective particle acceleration takes
place, at least for electrons. There are also arguments, like the
bilateral symmetry of the  X-ray synchrotron emission suggesting a roughly
uniform ambient magnetic field, that favor a type Ia origin for
G1.9+0.3. Finally, the distance estimate $d =8.5$~kpc is based on an analysis
of the absorption toward G1.9+0.3.

During the survey of the inner Galaxy by H.E.S.S. in very high energy
$\gamma$-rays, no emission was reported from the direction to G1.9+0.3
\citep{aha06}. Therefore, one can derive an upper limit at the
level of 2\% of the Crab flux above 200 GeV.

For the purpose of a more general study of such an unusual object regarding its
nonthermal properties, it is of interest to describe it by a kinetic theory of
cosmic ray (CR) acceleration in SNRs, coupled with the gas dynamics of the
thermal plasma, as given by \citet{bek96} and \citet{bv97}. This model assumes
spherical symmetry, although the assumption is later relaxed.  Similar models 
on almost the same physical basis
have recently been developed by two other groups \citep{KangJ06,zp09}, whose
calculations very well confirm these earlier results.
 
The kinetic description allows a corresponding analysis of the nonthermal
evolution of an SNR at a very early phase. This assumes that the plasma
  physics underlying especially the temporal dependence of the magnetic field
  amplification process can be extrapolated to such an early evolutionary
  phase, where the dynamical behavior of the ejecta plays an essential role.
Such a study is reported here. It is combined with a discussion about the
influence of the assumption of a smaller distance on the TeV \gr flux.

\section{Model}
Following \citet{rey08} it is assumed that G1.9+0.3 is a Type Ia supernova
(Type Ia SN) which expands into a uniform interstellar medium (ISM). 
Specifically, the object is assumed to eject a Chandrasekhar
mass $M_\mathrm{ej}=1.4 M_{\odot}$ with a total hydrodynamic explosion energy
$E_\mathrm{SN}= 10^{51}$~erg. During an initial period, the ejecta material
has
a broad distribution in velocity $v$. The fastest part of these ejecta is
described by a power law $d M_\mathrm{ej}/ d v\propto v^{2-k}$
with $k=7$ \citep[e.g.,][]{chev82}.

The ISM mass density $\rho_0=1.4m_\mathrm{p}N_\mathrm{H}$,
which is usually characterized by the hydrogen number density $N_\mathrm{H}$,
is an important parameter which strongly influences the expected SNR dynamics
and nonthermal emission; here $m_\mathrm{p}$ denotes the proton mass.

Following \citet{rey08} also a distance $d=8.5$~kpc is adopted for the main
part of
the paper. The observed shock size $R_\mathrm{s}=2$~pc and shock speed
$V_\mathrm{s}=14,000$~km~s$^{-1}$ are then used to determine the SNR age
$t_\mathrm{SN}$ and the ISM number density $N_\mathrm{H}$ for the given source
distance $d$.

As reviewed earlier \citep{voelk04, ber05, ber08} and elaborated most recently
in detail in \citet{bkv09}, the key parameters of the theoretical model (proton
injection flux density, given by a constant injection parameter $\eta \ll 1$
times the thermal particle flux density into the shock, electron-proton ratio
below the synchrotron cooling range, later denoted as $K_\mathrm{ep}$, and
magnetic field amplification) can be estimated in a semi-empirical way from a
fit of the theoretical solution to the observed synchrotron emission spectrum
if the characteristics of the initial explosion and
the relevant astronomical parameters are known. For this purpose, the
nonlinear aspects of the kinetic description are required.

Of the above-mentioned processes, the magnetic field amplification
\citep{bell04} is the least well understood. It is connected with the strong
nonlinear excitation of magnetic field fluctuations in the shock precursor by
the accelerating energetic particles
\citep{mckenzie82,lucek00,belllucek01,bell04}. Therefore, it is assumed here
that these fluctuations lead to Bohm diffusion of the energetic particles in
this amplified field. Such a bootstrap mechanism can be approximately justified
by the results of recent particle simulations \citep{rev08}. In addition, the
amplification process is strongly dissipative, as shown by hydromagnetic and
kinetic simulations
\citep{bell04,zirakashvili08,rev08,niemiec08,riquelme09,ohira09}.  Therefore,
it must be accompanied by strong gas heating within the precursor region
$r>R_\mathrm{s}$ due to wave dissipation, adopted in the present model in the
form $(\partial e_\mathrm{g}/\partial t)_\mathrm{diss} = -\alpha_\mathrm{H}
c_\mathrm{A} \partial P_\mathrm{c}/\partial r$ with $\alpha_\mathrm{H} =1$
\citep{bek96,bv97}, where $e_\mathrm{g}$ is the gas thermal energy density,
  $P_\mathrm{c}$ denotes the energetic particle pressure, and where, in a
second bootstrap mechanism, $c_\mathrm{A}=B(4\pi\rho)^{-1/2}$ is the Alfv\'en
velocity
in the amplified field $B$ (see below). The value $\alpha _\mathrm{H}=1$
corresponds to the assumption that the Alfv\'{e}n wave field excited within the
precursor reaches amplitudes which are very much smaller than the maximal
amplitudes which could be reached if the whole work $-c_\mathrm{A} \partial
P_\mathrm{c}/\partial r$ done by CRs went into wave excitation. In such a case,
this work goes almost completely into gas heating due to the wave damping. The
gas thermal pressure just ahead of subshock is in this case considerably larger
than the pressure of the magnetic field \citep{ber08}. Therefore, the subshock
can be treated approximately as a pure gas shock. This approximation will later
be re-examined through the approximate inclusion of the amplified field and its
associated turbulent gas motions in the subshock dynamics.

The magnitude of field amplification in all young SNRs is such that a
non-negligible fraction of the shock energy $\rho_0V_\mathrm{s}^2$ is converted
into magnetic field energy \citep[e.g.][]{ber08}. In fact, a time dependent,
amplified upstream magnetic field strength
\begin{equation}
B_0(t)=B_0(t_\mathrm{SN})[V_\mathrm{s}(t)/V_\mathrm{s}(t_\mathrm{SN})]^{\delta}
\label{B0}
\end{equation}
\noindent is used here, where the theory parameter $B_0(t_\mathrm{SN})$ is the
rms field strength at the present epoch $t_\mathrm{SN}$ and is estimated by a
comparison of the theoretically calculated synchrotron spectrum with the
observed one (see below). Such a form of the time dependence of the amplified
magnetic field with $\delta \approx 1$ is consistent with the interior field
strengths estimated from observational results for a number of young SNRs
\citep[e.g.][]{vbk05}. (Since \citet{bell04} even estimated a dependence
$B_0(t) \propto V_\mathrm{s}^{3/2}(t)$ for the field amplification due to the
nonresonant streaming instability alone, also the case of
$\delta =3/2$ will be examined here.  The radial dependence of the
rms. magnetic field strength $B(r,t)$ in the shock precursor is then modeled
by $B(r,t)=B_{0}(t) \rho(r,t)/\rho_0$, where $\rho_0$ is the far upstream
(interstellar) density. In the overall conservation relations for momentum and
energy of the system, the magnetic field strength in the upstream ISM
is $B_\mathrm{ISM} < B_0$.  Thus, $B_{0}/B_\mathrm{ISM}$ is the field
amplification factor by the accelerating energetic particles alone. An analysis
performed for a number of young SNRs shows \citep[e.g.,][]{ber08} that the field
strength $B_0$, required to fit the observed synchrotron spectrum, is well
within the range expected from theoretical estimates
\citep[e.g.,][]{bell04,pelletier06}.

The observed X-ray morphology of SNR G1.9+0.3 agrees with the theoretical
expectations regarding the morphology of ion injection and the corresponding
morphology of magnetic field amplification for a Type Ia SN \citep{vbk03}. It
is therefore consistent with a correction for the spherically symmetric
solution by a renormalization factor $f_\mathrm{re} \approx 0.2$ of the energy
density of nuclear particles, like in the case of SN~1006 \citep{bkv09}.  In
this case, the electron-proton ratio $K_\mathrm{ep}$, calculated under the
assumption of spherical symmetry, should be increased by a factor
$1/f_\mathrm{re}$.

\section{Results}

\begin{figure}
\plotone{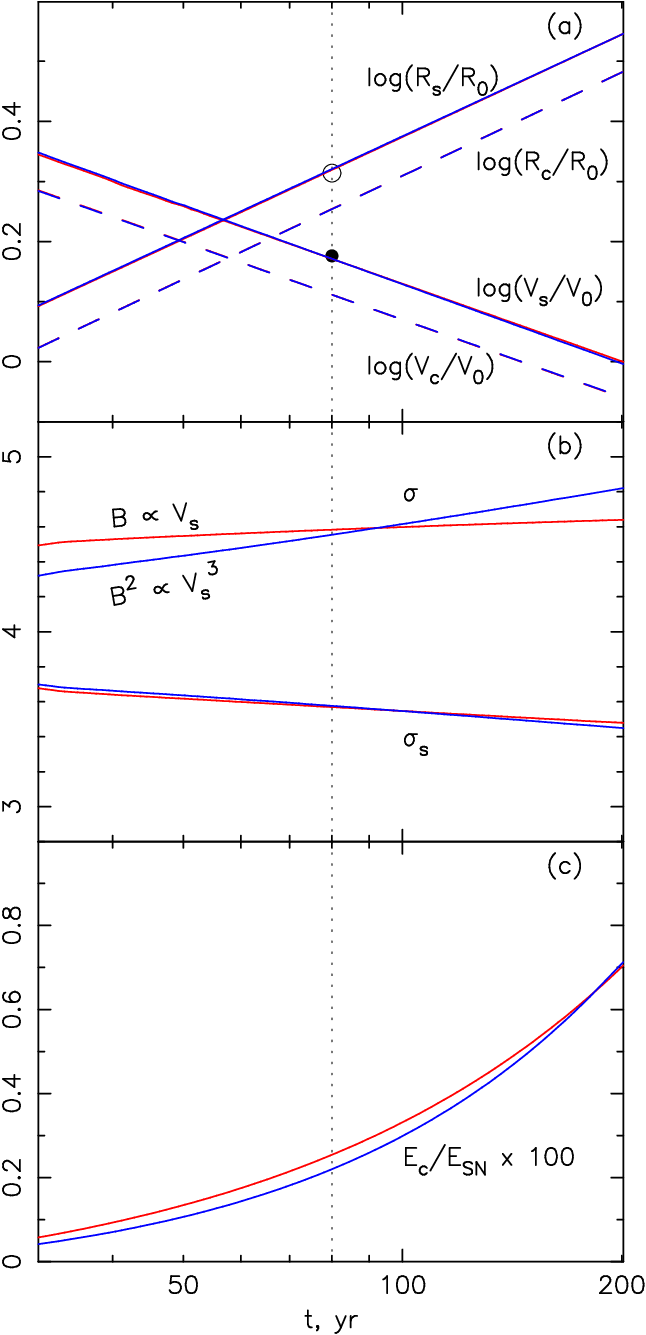} 
\figcaption{Shock (CD,
    (dashed lines)) radii $R_\mathrm{s}$ ($R_\mathrm{c}$) and shock
    (CD, (dashed lines)) speeds $V_\mathrm{s}$
    ($V_\mathrm{c}$) in units of $R_0=1$~ pc and $V_0 = 10^4$ km~s$^{-1}$ (a),
    total shock $\sigma$ and thermal subshock $\sigma_\mathrm{s}$ compression
    ratios (b), and total energy contents of accelerated CRs, $E_\mathrm{c}$
    (c), as functions of time in years. The red colored curves imply an
    amplified magnetic field strength $B\propto V_\mathrm{s}(t)$, whereas the
    blue colored curves correspond to $B \propto V_\mathrm{s}^{3/2}(t)$,
    cf. Equation (1); the difference between these two cases is not
distinguishable
    in Figure \ref{f1}(a). The dotted vertical line marks the current
epoch,
    $t_\mathrm{SN} = 80$~yr. The observed mean size {\it (open circle)} and
    speed {\it (filled circle)} of the shock, as determined by X-ray
    measurements \citep{rey08}, are shown as well.
\label{f1}}
\end{figure}

The calculated evolution of the gas dynamical variables of G1.9+0.3 is shown in
Figure~\ref{f1}. The observed shock radius $R_\mathrm{s}$ and shock speed
$V_\mathrm{s}$ are fitted at the age $t_\mathrm{SN} = 80$ yr and for an ISM
hydrogen number density $N_\mathrm{H}=0.018$ cm$^{-3}$ (Figure~\ref{f1}a). 
Since
the remnant is in the free expansion phase, it is approximately
consistent with an analytical self-similar solution $R_\mathrm{s}\propto
t^{4/7}$
\citep{chev82}. Note that an explosion model with an exponential ejecta
velocity profile gives slightly larger values of the age, $t_\mathrm{SN}= 100$
yr, and of the ISM density, corresponding to a hydrogen number density of
about $N_\mathrm{H}\approx 0.03$ cm$^{-3}$, for the same assumed distance of
$d=8.5$~kpc \citep{rey08}.

The calculated radius $R_\mathrm{c}$ of the contact discontinuity (CD) and the
CD speed $V_\mathrm{c}$ are also shown in Figure \ref{f1}(a). One can see that
the
ratio $R_\mathrm{c}/R_\mathrm{s}$ is rather small. At the current epoch
$R_\mathrm{c}/R_\mathrm{s}\approx 0.9$.

%-----------------------------------------------------------------Fig.1----
% ----------------------------------------------------------------------------

To obtain a good fit for the observed synchrotron spectrum (see below) first of
all a proton injection rate $\eta= 10^{-3}$ is required. It leads to a
nonlinear modification of the shock which, at the current age of
  $t=80$~yr, has a total compression ratio $\sigma \approx 4.6$ and a subshock
  compression ratio $\sigma_\mathrm{s} \approx 3.6$ (Figure \ref{f1}(b)). In
  addition, an electron-proton ratio $K_\mathrm{ep} \approx 5 \times 10^{-3}$
  and an upstream amplified magnetic field strength at the current epoch
  $B_0(t_\mathrm{SN}) \approx 100$~$\mu$G are required. This implies a
  downstream magnetic field strength of $B_\mathrm{d} \approx 460$~$\mu$G.

  All the above quantities are almost the same for the two different time
  dependences of the magnetic field $B_0(t)$ within the evolutionary period
  $t=t_\mathrm{SN} \pm 100$~yr. The only difference between these cases is a
  slightly more rapid increase of the shock modification (characterized by the
  shock compression ratio $\sigma(t)$) in the case $\delta = 3/2$ compared with
  the case $\delta = 1$ due to the increase of the Alfv\'enic Mach number
  $M_\mathrm{A}\propto V_\mathrm{s}/B_0\propto V_\mathrm{s}^{1-\delta}\propto
  t^{3(\delta-1)/7}$.

  To explain these results it is noted that, as in other, similar cases of
  strong, modified shocks, the existing measurements permit an estimate of
the three  parameters of the theoretical model.
This takes into account
  the following influence of the parameters on the synchrotron spectrum. (1)
  Since the high-energy electrons undergo strong synchrotron losses and, since
  during the acceleration process they dominate the nonthermal electron
  pressure $P_\mathrm{c}^\mathrm{e}$, the flux of X-ray synchrotron emission
  which they produce $\nu S_{\nu}$ is approximately proportional to the total
  energy flux of nonthermal electrons $F_\mathrm{e}\propto
  P_\mathrm{c}^\mathrm{e}V_\mathrm{s}R_\mathrm{s}^2\propto
  K_\mathrm{ep}P_\mathrm{c} V_\mathrm{s}R_\mathrm{s}^2$, and is only weakly
  sensitive to the magnetic field strength $B_0$. The spectrum of high-energy
  protons $N(p)$, which give the main contribution to the total CR pressure
  $P_\mathrm{c}\sim \rho V_\mathrm{s}^2$, is only weakly sensitive to the
  injection parameter $\eta$ in the case of a modified shock. Therefore, the
  fit of the {\it observed X-ray flux} mainly determines the value
  $K_\mathrm{ep}$ of the electron:proton ratio. The spectrum of accelerated
  electrons is calculated in absolute numbers. It is expressed in terms of
  $K_\mathrm{ep}$, due to the presumably dominant dynamical role of
  protons. (2) Values $\alpha >0.5 $ of the radio spectral index $\alpha =
  -d\ln S_{\nu}/d\ln \nu$, as observed in young SNRs, require a modified shock
  with $\sigma_\mathrm{s}<4<\sigma$ (This also implies a curved electron
  spectrum that hardens toward high frequencies). The value of $\alpha$ is
  mainly determined by the subshock compression ratio $\sigma_\mathrm{s}$,
  which in turn is determined mainly by the proton injection rate
  $\eta$. Therefore the fit of the {\it measured spectral shape} of the radio
  synchrotron emission gives mainly the required value of the injection
  parameter $\eta$. (3) Finally, since the radio emission flux value
  $S_{\nu}\propto K_\mathrm{ep}\nu^{-\alpha}B_\mathrm{d}^{\alpha +1}$ is
strongly
  dependent upon the magnetic field strength, its value $B_\mathrm{d}$ is
  derived from the fit to the {\it observed amplitude} of the radio spectrum.
  Thus, three measured characteristics of the synchrotron
    spectrum --- X-ray flux, shape, and amplitude of the radio emission --- make
it
    possible to obtain an estimate of the three relevant ``theory parameters''
    $B_\mathrm{d}$, $K_\mathrm{ep}$ and $\eta$ even though this is not a simple
    three-step procedure but an iterative procedure, minimizing the
combined
    $\chi^2$-value \citep[see][for details]{bkv09}.  

    Note that the magnetic field value $B_\mathrm{d}$ can also be estimated by
    another, independent method, namely from a fit of the {\it observed spatial
      fine structure} of the X-ray emission. In all the cases where such
    measurements exist, both methods give consistent values of
    $B_\mathrm{d}$ \citep[e.g.,][]{vbk05}.  Unfortunately, the fine structure of
    the X-ray emission is not determined yet for G1.9+0.3.  Therefore, this
      consistency check cannot be made at present.

    The uncertainties of the estimated values of $\eta$, $B_\mathrm{d}$, and
    $K_\mathrm{ep}$ depend upon the quality of the measurements of the
    synchrotron spectrum and can be rather small, as it was recently
    demonstrated for the case of SN~1006 \citep{bkv09}. In the case under
    consideration it is about 20\% for $B_\mathrm{d}$ and $K_\mathrm{ep}$ and 30\% for
 $\eta$.

It should also be noted that due to the very low age of the SNR and the low
ISM density the expected thermal X-ray emission is far below the observed X-ray
flux, which is therefore completely dominated by the nonthermal component. 

The required proton injection rate $\eta= 10^{-3}$ is considerably higher than
the critical value $\eta_{*}\approx 10^{-4}$, which separates a nonlinearly
modified shock with $\eta >\eta_{*}$ from an unmodified state,  
resulting from very low injection rates $\eta < \eta_{*}$
\citep[see][Eq.38]{berel99}.  The relatively weak shock modification is the
result of the very large magnetic field $B_0$, that leads to strong gas heating
within the precursor region $r>R_\mathrm{s}$. 
 
  In order to check the sensitivity of the results to the adopted value of the
  parameter $\alpha_\mathrm{H}$, calculations with $0.02\leq \alpha_\mathrm{H}
  <1$ were performed.  It turned out that even for $\alpha_\mathrm{H}=0.02$ the
  shock properties are not very far from the case $\alpha_\mathrm{H} =1$: the
  required far upstream magnetic field value is $B_0=125$~$\mu$G, and the shock
  compression ratios are $\sigma=6.3$ and $\sigma_s=3.99$.

  It is noted here that the results of \citet{vladimirov08} suggest a
  stronger sensitivity of the shock properties to the value of
  $\alpha_\mathrm{H}$.  The reason is that in the considerations of these
  authors the parameter $\alpha_\mathrm{H}$ determines not only the gas heating
  through the term $-\alpha_\mathrm{H} c_\mathrm{A} \partial P_c/\partial r$,
but also
  the upstream magnetic field generation through a complementary term
  $(\alpha_\mathrm{H}-1) c_\mathrm{A} \partial P_\mathrm{c}/\partial r$. Such
  an approach implies that magnetic field amplification takes place only due to
  resonant excitation of Alfv\'{e}n waves and, in addition, according to such a
  quasi-linear expression. In contrast, the present  model
  allows that the magnetic field is also amplified nonresonantly
  \citep{bell04,pelletier06}.  Secondly, the determination of the amplified
  field $B_0(t_\mathrm{SN})$ in the present semi-empirical model uses the
  synchrotron observations of the source. For both these reasons, the magnetic
  field amplification and the gas heating are not connected by a simple
  relation.

\subsection{Subshock dynamics with amplified B-field}

Following \citet{mckenzie82} and \citet{vladimirov08}, an Alfv\'enic connection
${\bf w}={\bf B}(4\pi \rho)^{-1/2}$ between the magnetic field fluctuation
vector ${\bf B}$ and the fluctuation vector ${\bf w}$ of the mass velocity is
assumed. This implies an approximately incompressible plasma turbulence with a
locally homogeneous mass density $\rho$, where the total (mean square magnetic
field plus plasma turbulent) pressure is given by $P_\mathrm{turb}= ( {\bf
  B})^2/(8\pi)$, and the total (mean square magnetic field plus plasma
turbulent) energy flux density equals $F_\mathrm{turb} = 3uP_\mathrm{turb}$;
here $u$ denotes the shock-normal mean mass velocity in the shock frame. These
normal components of the momentum and energy flux densities, immediately
upstream and downstream of the subshock were included in the Rankine--Hugoniot
conditions for the subshock. The latter is approximated as a
locally plane, normal shock wave.  (In the precursor region, $({\bf B})^2 = B^2$
is the mean square strength of the amplified magnetic field $B$, introduced in
the previous section and assumed here to be isotropically distributed with
Gaussian statistics; the field strength downstream of the subshock is taken to
be $B_2=\sigma_\mathrm{s} B_1$, where $B_1$ is the field strength upstream of
the subshock.)  Note that the above expression for the total turbulent energy
flux density $F_\mathrm{turb}$ in the downstream region differs somewhat from
the expression used by \citet{caprioli08}. Since the consideration of these
authors is based on the transmission and reflection of small-amplitude Alfv\'en
waves at purely parallel subshock, it is believed here that this linear
treatment is not applicable to the actual case of a strongly perturbed and
amplified magnetic field (see also the arguments of \citet{vladimirov08}).

It is clear that the approximations introduced above do not exactly
  describe the true physical situation thatalso  contains the nonresonantly
  unstable modes of the Bell instability \citep{bell04}, because these
  transverse modes will in their nonlinear evolution also develop compressible
  elements \citep{bell04,zirakashvili08}. In addition, also the
  acoustic modes \citep{dorfi84,drury84,ber86,malkov06}
 will contribute. Their influence on the strength of the subshock
  remains to be evaluated. However it is believed that the present description
  gives a roughly correct estimate of the subshock effects of at least the
  incompressible part of the fluctuation fields produced by the accelerating
  particles.

In  the adopted approximation $\sigma=4.55$,
$\sigma_\mathrm{s}=3.6$  for $\alpha_\mathrm{H} = 1$, 
very close to the previous
case, where the  turbulent momentum
and energy fluxes were ignored in the subshock conservation
relations. The effect is somewhat larger for the smallest value of the
parameter $\alpha_\mathrm{H}=0.02$: it leads to a decrease of $\sigma$ from
$\sigma=6.3$
to $\sigma=5.9$ and to a decrease of the magnetic field strength from
$B_0=125$~$\mu$G to $B_0=120$~$\mu$G, which is not a large effect either.

It should also be noted that the assumption of considerable gas heating due to
wave dissipation, corresponding to $\alpha_\mathrm{H} =0.5-1$, is consistent
with the
numerical modeling of the nonresonant wave excitation
\citep{bell04,zirakashvili08}. Such dissipation should
operate in a similar way for the resonant Alfv\'{e}n mode instability. It is
therefore concluded that insignificant gas heating, which occurs for
$\alpha_\mathrm{H} \ll 1$ within the present formalism, is an unrealistic
assumption.

\subsection{Charged particle and $\gamma$-ray spectra}

With the renormalization $f_\mathrm{re} = 0.2$, the nuclear CRs inside G1.9+0.3
SNR contain (Figure \ref{f1}(c))
\begin{equation}
  E_\mathrm{c}\approx 0.0025E_\mathrm{SN} \approx 3\times10^{48}~\mbox{erg}.
\end{equation}

\begin{figure} 
\plotone{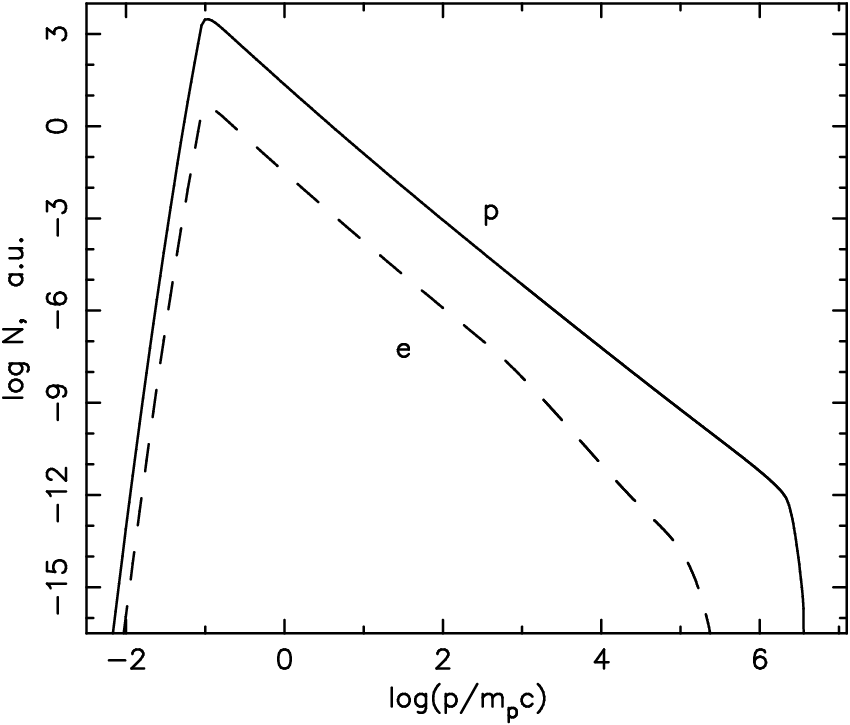}
\caption{Spatially integrated CR spectrum as function of particle momentum. 
  Solid and dashed lines correspond to protons and electrons,
  respectively.
\label{f2}}
\end{figure}

The volume-integrated (or overall) CR spectrum
 \begin{equation}
N(p,t)=16\pi^2p^2 \int_0^{\infty}dr r^2 f(r,p,t)  
\label{N}
\end{equation} 
has, for the case of protons, almost a pure power-law form $N\propto
p^{-\gamma}$ over a wide momentum range from $0.1m_\mathrm{p}c$ up to the
cutoff momentum $p_\mathrm{max}\approx 3\times 10^6m_\mathrm{p}c$
(Figure \ref{f2}). This value $p_\mathrm{max}\propto
R_\mathrm{s}V_\mathrm{s}B_0$
is limited mainly by the finite size and speed of the shock, its deceleration
and the adiabatic cooling effect in the downstream region \citep[see][for
  details]{ber96}. As pointed out above, particle diffusion is approximated by
Bohm diffusion in the amplified magnetic field $B$, cf. Equation (1). It is
important
to note that the calculated value of $p_\mathrm{max}$ is therefore an
upper limit, because it is assumed that up to the cutoff all particles ``see''
the amplified field everywhere.

Consequently, G1.9+0.3 represents the youngest SNR where the accelerated proton
spectrum extends up the so-called knee energy. Such a maximum proton energy
appears indeed required to describe the overall CR spectrum for energies up to
$10^{17}$~eV \citep{bv07}.

The shape of the overall electron spectrum $N_\mathrm{e}(p)$ deviates from that
of the proton spectrum $N(p)$ at high momenta $p>p_\mathrm{l}\sim
10^3m_\mathrm{p}c$ on account of the synchrotron losses during the electron
residence time in the downstream region (Figure \ref{f2}). Within the momentum
range $p_\mathrm{l}< p < p_\mathrm{max}^\mathrm{e}$, the electron spectrum is
considerably steeper, $N_\mathrm{e}\propto p^{-3}$, due to synchrotron losses
taking place in the downstream region after the acceleration at the shock
front.  The maximum electron momentum $p_\mathrm{max}^\mathrm{e}\approx
10^5m_\mathrm{p}c$ corresponds closely to the result obtained
by equating the synchrotron loss time and the acceleration time.

\begin{figure} 
\plotone{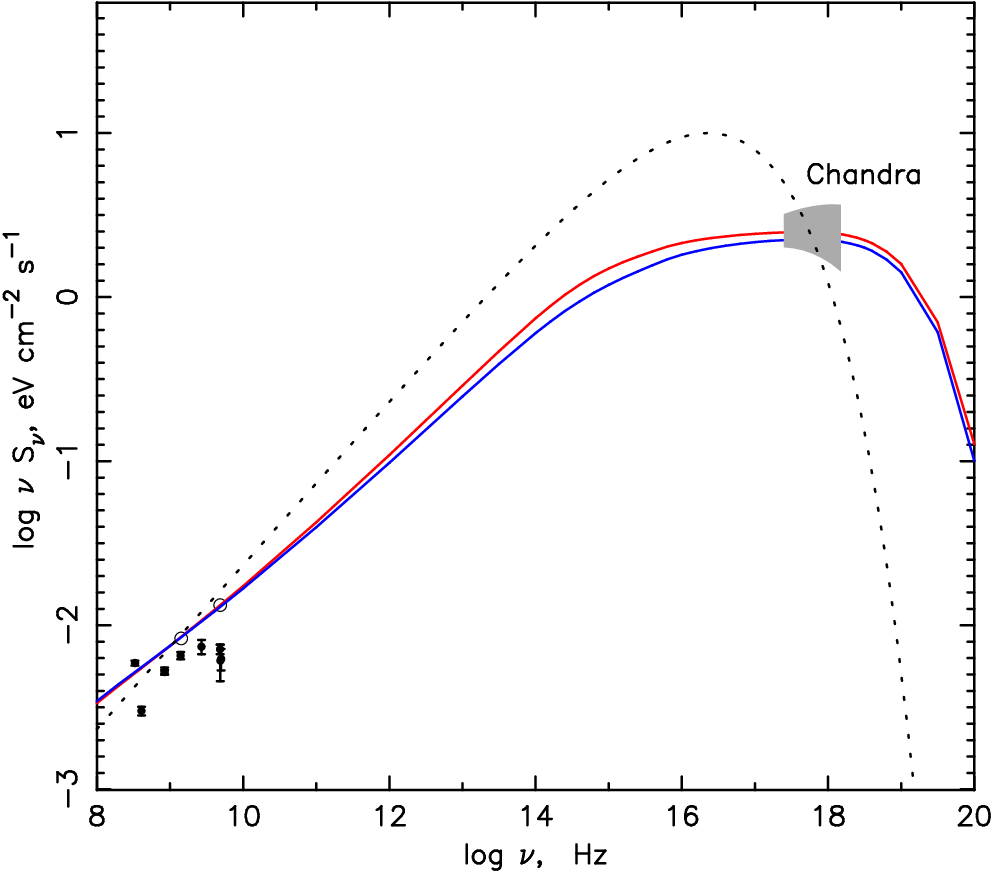} 

\figcaption{Spatially integrated synchrotron SED as a
  function of frequency. The dotted line corresponds to the test particle
  limit (leptonic scenario). Fluxes of the X-ray emission observed by {\em
  Chandra} \citep{rey08} and the radio emission compiled by \citet{green08} are
  also shown. The solid lines are fitted to the most recent VLA radio
  data shown by the open circles, reported by \citet{green08}.  The 
  line colors have the same meaning as in Figure \ref{f1}.
\label{f3}}
\end{figure}

Figure \ref{f3} illustrates the consistency of the synchrotron spectrum,
calculated for the above-mentioned best set of parameters with the observed
spatially integrated spectra.
% 
%------------------------------------------------------------------------fig.2-
%------------------------------------------------------------------------------

As mentioned above, values $\alpha >0.5$ of the radio spectral index $\alpha =
-d\ln S_{\nu}/d\ln \nu$, as observed in young SNRs, require a curved electron
spectrum that hardens toward higher energies, as predicted by nonlinear shock
acceleration theory. To have $\alpha=0.62$ in the radio range, as observed for
G1.9+0.3 \citep{green08}, requires efficient CR acceleration with a proton
injection rate $\eta= 10^{-3}$ which leads to the required shock modification,
and also leads to the high magnetic field value above.  As it is clear from
Figure \ref{f3}, a good fit of the observed X-ray energy flux can only be
achieved
due to the softening of the synchrotron spectrum for $\nu \gsim 10^{15}$~Hz,
which is due to the strong synchrotron losses of electrons with momenta
$p>p_\mathrm{l}\approx 700m_\mathrm{p}c$. Using the known dependence
$p_\mathrm{l}\propto B^{-2}_\mathrm{d}t^{-1}$, at $t=t_\mathrm{SN}$ one can
immediately estimate the required value of the interior magnetic field value
$B_\mathrm{d}\approx 500$~$\mu$G,
consistent with the above determination from the radio spectrum. The calculated
hard power-law X-ray spectrum continues almost up to 50~keV, making this source
in principle attractive to be observed with {\em Suzaku} and {\em INTEGRAL} and
the future Astro-H X-ray instrument.

A very important question for every young SNR is whether the existing data
indeed unavoidably require efficient proton acceleration accompanied by strong
magnetic field amplification. In order to explore the alternative possibility,
Figure \ref{f3} presents in the dotted curve a synchrotron spectrum which
corresponds to a hypothetical leptonic scenario with a proton injection rate so
small ($\eta\ll10^{-4}$) that the accelerated nuclear CRs do not produce any
significant shock modification and therefore also no magnetic field
amplification. This corresponds to the test particle limit, when the
distribution function of shock accelerated electrons has the form
\begin{equation}
f_\mathrm{e}= A p^{-4}\exp(-p/p_\mathrm{max}),
\label{fe}
\end{equation}
where the amplitude $A$ and the value of the cutoff momentum $p_\mathrm{max}$
are determined by the fit to the observed synchrotron spectrum for a given
interior magnetic field value $B_\mathrm{d}$. Since magnetic field
amplification is not expected in this case, the downstream magnetic field cannot
be larger than the MHD-compressed ISM field $B_\mathrm{ISM}\approx
5$~$\mu$G. The maximal possible downstream field $B_\mathrm{d} \approx
20$~$\mu$G is adopted which corresponds to the minimal number of accelerating
electrons, and therefore the \gr emission produced by these electrons is also
minimal. The synchrotron spectrum for the leptonic test particle scenario
in Figure \ref{f3} corresponds to the maximal electron energy
$\epsilon_\mathrm{max}^\mathrm{e}=p_\mathrm{max}^\mathrm{e}c=6$~TeV, determined
by the {\em Chandra} observation. There are two differences in the synchrotron
spectra, corresponding to these two scenarios. The high-injection scenario
leads to a soft radio spectrum $S_{\nu}\propto \nu^{-\alpha}$ with power law
index $\alpha=0.62$, whereas in the test particle case $\alpha=0.5$. On the
other hand, the two spectra behave essentially differently at X-ray frequencies
$\nu\gsim 10^{18}$. This demonstrates that only in the high-injection case with
its high, amplified magnetic field value 
$B_\mathrm{d}\approx 460$~$\mu$G the spectrum $S_{\nu}(\nu)$ has a smooth
cutoff, consistent with the observations (see Figure \ref{f3}). In the test
particle case the spectrum $S_{\nu}(\nu)$ has too sharp a cutoff to be
consistent with the observations.

If G1.9+0.3 was indeed a Type Ia SN, then the
  explosion parameters $E_\mathrm{SN}$, $M_\mathrm{ej}$ and $k$ are known. 
  Assuming the value $d=8.5$~kpc for the source distance and using the age
$t_\mathrm{SN}$ and the ambient gas number density $N_\mathrm{H}$ from fits to
the observed
astronomical parameters size and expansion rate, and also using the
``theory parameter'' values $\eta$, $K_\mathrm{ep}$ and $B_0$, estimated from
the fit
to the synchrotron spectrum, one can predict the  $\gamma$-ray
flux for the assumed source distance $d$.

%------------------------------------------------------------------fig.3---
%--------------------------------------------------------------------------
In Figure \ref{f4}, the calculated $\gamma$-ray spectral energy 
distributions (SEDs) due to $\pi^0$-decay and inverse
Compton (IC) collisions are presented 
for the source distance $d=8.5$~kpc
together with the sensitivities of the
{\em Fermi} and H.E.S.S. instruments. For the modified shock, consistent with
  the observed synchrotron emission, the expected total TeV $\gamma$-ray SED
is  $\epsilon_{\gamma}F_{\gamma}\approx 3 \times
  10^{-3}$~eV~cm$^{-2}$~s$^{-1}$. Such a flux is too low for an  H.E.S.S.
detection in $\sim 50$~hr by a factor of the order of 
  30. The $\pi^0$-decay flux is only about  6\% of the IC
\gr flux as a result of the low gas density. Since according to recent
estimates \citep{Porter06}, the Galactic interstellar optical and infrared
radiation fields in the inner Galaxy are considerably higher than previously
thought, for this region the calculation of the IC flux
was performed on the basis of those estimates. 
The higher radiation field leads to an increase of the IC gamma-ray flux at
energies $\epsilon_{\gamma}< 1$~TeV by an order of magnitude compared with
  a standard interstellar radiation field in the solar neighborhood
  \citep[e.g.][]{dav94}.  Nevertheless, as indicated above, the expected TeV
  emission flux is still far below the H.E.S.S. sensitivity.

  As can be seen from Figure \ref{f4}, the TeV \gr flux expected in the
unmodified
  leptonic scenario considerably exceeds the H.E.S.S. sensitivity,
  corresponding to $\sim 50$~hr of observation time. Since the region of the
  Galactic center was already explored by H.E.S.S. for times of more than 100~hr
  without detection of G1.9+0.3, this  purely leptonic test particle
  scenario should be rejected as in all similar cases of Type Ia SNe
  \citep{vbk08}.

\subsection{Dependence on the assumed source distance}

It is, however, to be noted that the expected \gr flux is very sensitive to the
assumed source distance $d$. Therefore, SNR G1.9+0.3 could be a potential \gr
source if the actual distance was lower than 8.5~kpc. Qualitatively, the
dependence of the expected \gr flux on distance can be understood if one
takes into account that the $\pi^0$-decay \gr energy flux $\epsilon_{\gamma}
F_{\gamma} \propto M_\mathrm{sw}e_\mathrm{c}/d^2$ is proportional to the mass
of gas, $M_\mathrm{sw}=4\pi R_\mathrm{s}^3\rho_0/3$, swept up by the SN shock,
and to the energy density $e_\mathrm{c}$ of the CRs producing \grs of given
energy. Since for high acceleration efficiency $e_\mathrm{c}$ is proportional
to the shock kinetic energy density $\rho_0V_\mathrm{s}^2$, one can write
\begin{equation}
\epsilon_{\gamma} F_{\gamma} \propto
N_\mathrm{H}^2V_\mathrm{s}^2R_\mathrm{s}^3/d^2.
\label{eq2}
\end{equation}
For fixed explosion energy $E_\mathrm{SN}$, the distance $d$ and ISM density
$N_\mathrm{H}$ are connected by the relation
\begin{equation}
N_\mathrm{H}\propto d^{-7},
\label{eq1}
\end{equation}
because in the free expansion phase the SNR radius $R_\mathrm{s}\propto d$ is
determined by the expression $R_\mathrm{s}\propto N_\mathrm{H}^{-1/7}t^{4/7}$
\citep{chev82}, where the SNR age $t\propto R_\mathrm{s}/V_\mathrm{s}$ is fixed
if the angular size and angular expansion speed are known as in our case of
G1.9+0.3. 

\begin{figure}
  \plotone{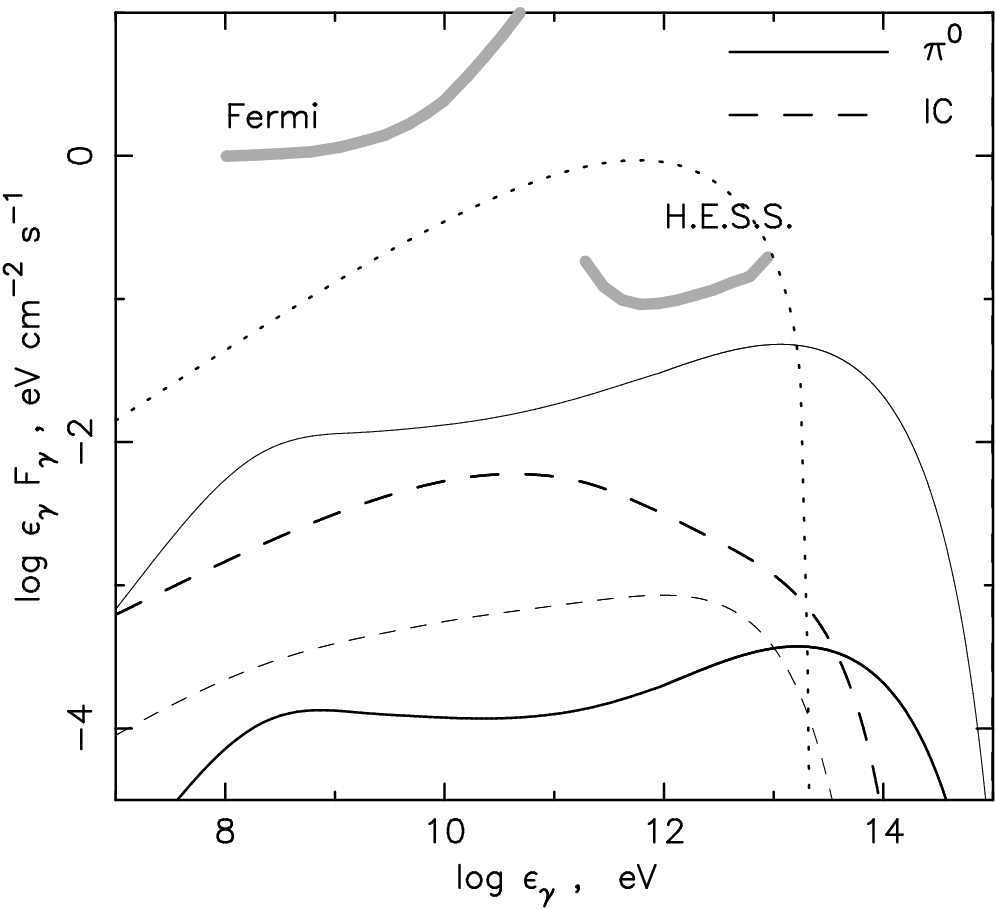} 

\figcaption{Integral $\pi^0$-decay (solid lines) and IC (dashed
    lines) $\gamma$-ray energy flux densities (SEDs) for the present epoch, as
    a function of $\gamma$-ray energy, for the two different source distances
    $d = 8.5$ kpc (thick lines) and $d = 5.6$ kpc (thin lines). The
    calculations are given for the case $\delta = 1$.  SEDs calculated for the
    case $\delta = 3/2$ coincide with the presented SEDs within 10\% accuracy.
    The thick dashed line represents the IC $\gamma$-ray energy flux,
    calculated for the recently re-estimated interstellar optical/infrared
    radiation background in the central region of the Galaxy \citep{Porter06}.
    The dotted line corresponds to the test particle limit (leptonic
    scenario). For comparison, the sensitivities of {\em Fermi} (for a $5
    \sigma$ detection in 1 yr of sky survey exposure with a background
    representative of the diffuse background near the galactic plane;
    \citet{atwood09}) and H.E.S.S. (for a $5 \sigma$ detection of the Crab
    Nebula with power-law differential photon index 2.6 in 50 hr at a zenith
    angle of 20$^{\circ}$; \citet{Funk2005}) are shown.
\label{f4}}
\end{figure}

Taking also into account that for a fixed angular expansion rate of the object
$V_\mathrm{s}\propto d$, one obtains
\begin{equation}
\epsilon_{\gamma} F_{\gamma} \propto d^{-11}.
\label{eq3}
\end{equation}
According to this relation a mere 30\% reduction of the source distance leads
to an increase of the expected \gr flux by a factor of 
more than 10. This is
illustrated in Figure \ref{f4}, where also \gr spectra are presented that were
calculated for the distance value $d=5.6$~kpc. In this case the shock velocity
and size could be fitted at the same age $t_\mathrm{SN} = 80$ yr and for an
ISM hydrogen number density $N_\mathrm{H}=0.2$ cm$^{-3}$. A similar fit for the
radio and X-ray data as in Figure \ref{f4} could be achieved with an
electron-proton ratio $K_\mathrm{ep} = 8 \times 10^{-4}$ and a downstream
magnetic field strength $B_\mathrm{d}=670$~$\mu$G.  It is clear that G1.9+0.3
could be visible in TeV \grs by future instruments like the Cherenkov 
Telescope Array (CTA), if the actual
distance was not larger than $d=5.6$~kpc.

%-------------------------------------------------------------fig4--------
%--------------------------------------------------------------------------

\begin{figure}
%\epsscale{0.60}
  \plotone{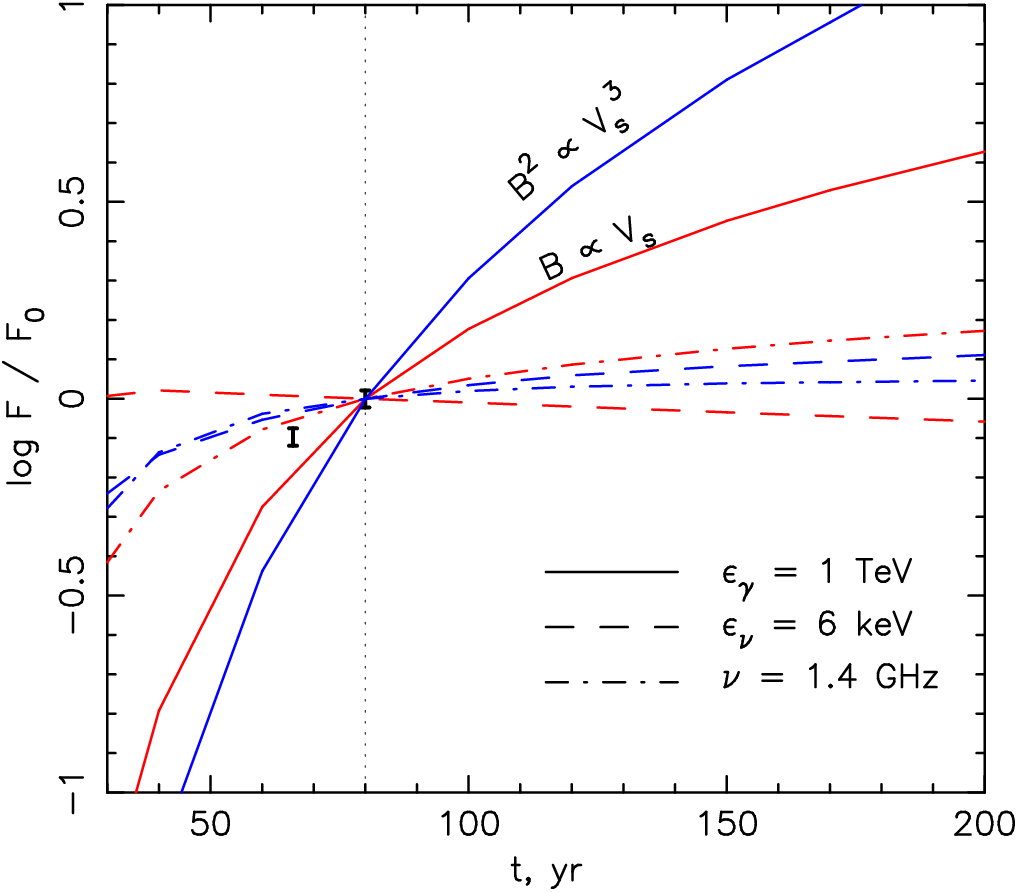} 
  \figcaption{Time dependence (in years) of the fluxes of the radio synchrotron
    emission at frequency $\nu=1.4$~GHz (dash-dotted lines), synchrotron
    X-ray emission with energy $\epsilon_{\nu}=6$~keV (dashed lines), and
    TeV energy \gr emission (solid lines). The fluxes are normalized to
    their values $F_0$ at the current epoch. Available data in radio
    \citep{green08} are shown as well. The line colors have the same
      meaning as in Figure \ref{f1}.
\label{f5}}
\end{figure}

The  theoretical, spatially integrated radio
synchrotron flux slowly increases with time, as can be seen in Figure \ref{f5},
essentially due to the rapidly increasing total number of accelerated
  electrons in the increasing SNR volume $\propto R_\mathrm{s}^3$.

The X-ray synchrotron flux is expected to be nearly
constant in time (Figure \ref{f5}). 
 This is mainly due to the strong synchrotron
cooling of the highest energy electrons which produce the X-ray synchrotron
emission.

The TeV \gr flux is expected to increase with time as well (Figure \ref{f5}),
mainly due to the increase of overall number of CRs with energy above 10~TeV.

%--------------------------------------------------------------fig.5
%------------------------------------------------------------------------------

\subsection{Renormalization of the nuclear particle spectra}

If the ambient interstellar magnetic field would be completely disordered on
  spatial scales smaller than the shock size, then efficient CR
  injection/acceleration would be expected across the whole shock surface. In
  such a case $f_\mathrm{re}\approx 1$ and the expected gamma-ray flux would be
  higher by a factor of $1/f_\mathrm{re}\approx 5$, whereas all the emission
  produced by CR electrons would remain the same since the normalization of the
  electron spectrum was done based on the observations. However, it is believed
  that the actual situation is opposite: the bilateral symmetry of the
  X-ray synchrotron emission suggests a roughly uniform ambient magnetic field
  on a parsec scale and therefore $f_\mathrm{re}\approx 0.2$ like in SN~1006.

\section{Conclusions}
The existing data for G1.9+0.3, when analyzed within the framework of the
nonlinear kinetic theory of CR production in SNRs described above, are
consistent with a type Ia explosion in a rarefied medium at a distance of
$d=8.5$~kpc, whose nuclear CR spectrum reaches the energy of the ``knee'' in
the observed Galactic CR spectrum at the present epoch. This conclusion also
concerns the derived strong magnetic field amplification. A test
  particle, purely leptonic gamma-ray scenario, is inconsistent with existing
TeV gamma-ray observations with the H.E.S.S. telescope array. However, the data
set is not complete enough to unequivocally determine the value of the source
distance. Since the expected \gr flux is very sensitive to the distance $d$,
$F_{\gamma} \propto d^{-11}$, a detection of the \gr flux from G1.9+0.3, as
improbable as it may be, would yield the distance. However, in the case when
G1.9+0.3 is located near the Galactic center ($d=8.5$~kpc), the expected TeV \gr
energy flux is so low, $\epsilon_{\gamma}F_{\gamma}\approx
5\times10^{-15}$~erg~cm$^{-2}$~s$^{-1}$, that it is not detectable with present
instruments. It is clear that for a distance that was not larger than
$d=5.6$~kpc G1.9+0.3 could be visible with future instruments like CTA which
can be assumed to have a sensitivity of $\sim 1$~mcrab at a few 100 GeV
\citep{bernloehr09}.

\acknowledgments

We are indebted to Dr. Stephen Reynolds for providing us the X-ray spectra for
G1.9+0.3 from {\em Chandra} in physical units. This work has been supported in
part
by the Russian Foundation for Basic Research (grants 07-02-00221,10-02-00154), 
Federal Agency of Science and Innovations (contract 02.740.11.0248), Program of
PRAS No. 16 and by the Leading Scientific Schools of Russia (project
3526.2010.2). EGB and LTK acknowledge the hospitality of the
Max-Planck-Institut f\"ur Kernphysik, where part of this work was carried out.

\end{document}